
\magnification=\magstep1
\baselineskip=12 true pt
\hsize=15 true cm
\vsize=22 true cm

\def\c{\centerline}
\def\zbb {\Gamma (Z\rightarrow b\bar b)}
\def\del {\delta^{\chi}_b (m_t)-\delta^{\chi}_b (0)}
\def\coeff {{\alpha\over {4\pi \sin ^2\theta_W}}}
\def\r {\rightarrow}
\def\zb {\Gamma^0 (Z\rightarrow b\bar b)}
\def\z {\Gamma^1 (Z\rightarrow b\bar b)}
\def\bl {\delta_b^{(t)}(m_t)}
\def\ssq {\sin ^2\theta _W}
\def\csq {\cos ^2\theta _W}

Recently, it has been shown [1-4] that the  electroweak  symmetry
of  the  Standard  Model may be broken dynamically by a $t\bar t$
condensate. This is referred to in the literature [5] as ``top-mode
Standard Model". The top quark, being much heavier than the other
known fermions (and lying close in the mass spectrum
to the electroweak scale  $v=246$
GeV), may, in this picture,  be  responsible for the breaking of
the $SU(3)_c\times SU(2)_L\times U(1)_Y$ to $SU(3)_c\times U(1)_{em}$.
It  has
been shown [2] that in this model, where
the   presence  of  a  four-fermion  interaction  of  the  form
$G(\bar\psi_L
t_R)(\bar  t_R\psi_L)$  induces  the symmetry breaking, the
bound-state spectrum  consists  of   three   massless   Nambu-Goldstone
bosons,  which  give masses to the massless gauge bosons, and  one
massive neutral scalar, which may be identified as the Higgs.

This   model   has several attractive features.  First,  the
naturalness problem arising in the elementary  scalar  sector  of
the  Standard Model can be isolated in the coupling constant $G$ once
for all. Second, no elementary scalar is necessary for  the
theory,  and  there  is  no  problem  regarding  the violation of the
unitarity bound in $WW\r WW$ scattering.
Third,  a  definite  relationship  can  be
established between the  top mass and  the   Higgs   masses  reducing
thereby (to  some  extent)  the  embarrasingly great laxity in the
choice of parameters of  the  otherwise  so  successful  Standard
Model. Finally,
there are physical examples of dynamical symmetry breaking at the
eV scale (BCS theory of superconductivity) and the MeV scale (the
breaking  of chiral symmetry for nucleons) and it would be aesthetically
satisfying if the mechanism should recur  again  at  this  higher
scale of energy.

Although the above model is elegant and economical in the sense that
it does not predict any new particle (even the Higgs scalar is  a
composite   object),  unfortunately the  top-quark mass  $m_t$ in
this model, as determined from the
renormalization  group  flow  of  the  coupling   constants,
appears to be untenable with the present experimental upper bound
of 190 GeV. To resolve this difficulty within the same framework,
  it  was  proposed  [6-8]
that  one  can  include an additional $SU(3)_c\times  SU(2)_L\times U(1)_Y$
invariant term in the Lagrangian, which is of the form
$$
G'(\bar\psi _{Li}^I(A^P_Q)_{IJ}t^J_R)(\bar t ^K_R
(A^Q_P)_{KM}\psi ^M_{Li}). \eqno(1)
$$
Here $G'$ is the coupling constant (of mass dimension $-2$)
for the new  interaction,  $i$
is the  $SU(2)_L$ index and $I$, $J$, $P$, $Q$  are the $SU(3)_c$
indices running from 1 to  3.  The  $A$  matrices  are  the  real
generators   of   $SU(3)$  {\it \'a  l\'a}  Okubo  which  we  find more
convenient for our problem than the usual Gell-Mann matrices [9].
The four-fermionic interaction being
nonrenormalizable  in  $3+1$  dimensions,  a  high  energy  cutoff
$\Lambda$  is  needed  for  the  regularization  of  this theory.
Effectively, this means that the theory ceases to be valid beyond
$\Lambda$.  For  simplicity,  we will use the same cutoff for
all four-fermionic operators.

In a theory with strong coupling,  one  can  use  the  perturbative
analysis  in  the  low-energy  limit  by introducing
auxiliary fields [10] in the action. Alternatively, one can write
down  a low-energy Lagrangian, which, at a high-energy scale, when
all  auxiliary  fields  are  integrated  out,  gives   back   the
Lagrangian  with four-fermionic interaction. For this, one has to
suitably define the different renormalization  constants,  taking
account  of  compositeness  [2].  Following the latter approach, we
will define the effective potential of our theory to be
$$
V=-\mu^2\phi^i\phi_i+m^2\chi^{iI}_J\chi^J_{iI}+a_1(\phi^i\phi_i)
^2+a_2(\chi^{iI}_J\chi^J_{iI}) ^2$$
$$+a_3(\chi^{iI}_J\chi^J_{iK}\chi^{jK}_M\chi^M_{jI})
+a_4(\chi^{iI}_J\chi^J_{jI}\chi^{jK}_M\chi^M_{iK})$$
$$+a_5(\phi^i\phi_i\chi^{jJ}_K\chi^K_{jJ})+a_6(\phi^i\phi_j
\chi^{jJ}_K\chi^K_{iJ})~~~~~~~~$$
$$~~~~~~~~~~~~~+g_t(\bar\psi_{Li}t_R\phi^i+\   {\rm    h.c.})\
+g'_t(\bar\psi
^I_{Li}(A^P_Q)_{IJ}t^J_R\chi^{iQ}_P+\ {\rm h.c.}). \eqno(2)
$$

It may be noted that  there  are  two composite
doublets  in  $V$,  namely,  $\phi$  and  $\chi$.  They have the same
isospin properties, but under $SU(3)_c$ the former is  a  singlet
while  the  latter  is an octet, which is denoted by the explicit
$SU(3)_c$ indices. The  field  $\phi$  arises  from  the  $SU(3)_c$ singlet
bilocal fermionic  field  $\bar\psi  _Lt_R$ while the $\chi$ field arises
from  the $SU(3)_c$ octet bilocal
 $\bar\psi_L^I(A^P_Q)_{IJ}t_R^J$. It  is   evident  that
$\phi_i^{\dag}\equiv     \phi^i$    and   $(\chi^I_{iJ})^{\dag}\equiv
\chi ^{iJ}_I$,  because  $(A^P_Q)^{\dag}=(A^P_Q)^T=(A^Q_P)$.  The
$\phi$  field  behaves  in  the  low energy limit as the Standard
Model Higgs doublet with $<\phi >_0=v$, while  the  $\chi$  field
has  zero  VEV since it carries color. The absence of a VEV for the
$\chi$  doublet  has  two  important  implications:  first,  it
guarantees  that  in  the  symmetry-broken phase, the SM relation
$v=\sqrt{-\mu^2/a_1}$ is preserved, and second, it allows  us  to
introduce  a gauge-invariant mass term of the form $m^2\chi^2$ in
the Lagrangian. The $a_i$ ($i$=1  to  6)  parameters  generically
denote  four-boson  couplings; the $SU(2)$ and $SU(3)$ indices of
$\chi$ can be contracted in three different  ways  to  produce  a
singlet.  Gauge-invariant  kinetic  terms for the $\phi$ and the
$\chi$ fields are also induced.

We can now solve numerically a set of $\beta$-functions corresponding
to  the parameters $a_i$, $g_t$ and $g'_t$, and find the infrared
quasi-fixed point solution to be [7]
$$a_1=0.18,\ \ a_5=0.18,\ \ a_6=0.42 \eqno(3)$$
$${9\over 4}g_t^2+2{g'_t}^2={9\over 4}(g_t)^2_{BHL}. \eqno(4)$$
We have not shown the other couplings at  the  quasi-fixed  point
because they are not relevant for our analysis, but they turn out
to be small and positive. It is to be noted that  the  parameters
are  not exactly those shown in eq. (2) but suitably renormalized
ones, and we use the same symbol only for brevity. The  value  of
$a_1$  gives  $m_H=209$ GeV. The suffix BHL indicates the results in
ref. 2, and
it is evident from eq. (4) that $m_t$ is a free parameter of  the
theory,  being  always  less than $(m_t)_{BHL}$. Therefore, it is
not possible to predict $m_t$ in this model. In the succeeding analysis,
we take some  phenomenologically plausible values for $m_t$.

The mass of $\chi$ will be an
important theme in
our discussion. We immediately note that though the field is an
auxiliary one arising as a composite of two  spinor  fields  in  an
$SU(3)_c$  octet combination, it is not possible to predict the
masses as was done for the ``Higgs scalar"  in  ref.  2,  because
here the strong interaction plays a nontrivial part and the 1/$N$
approximation is not  valid. Not being determinable  from  the
renormalization  group  equations, $m^2$ remains a free parameter
of the theory. When the symmetry is broken, two more terms of  the
form $\textstyle{1\over 2} a_5v^2$ and $\textstyle{1\over 2} a_6v^2$
contribute to $m_{\chi}^2$, the second term contributing only for
the neutral field, so that
$$m_{\chi^0}^2-m_{\chi^{+}}^2={1\over 2}a_6 v^2. \eqno(5)$$

These new bosons can have  profound consequences through  various
one-loop effects which are experimentally observable, and  we
can place lower bounds on their masses.
One notes  that  the  interaction  in  the  minimal
condensate  scheme  is  confined  only to the third generation of
quarks. The other quark generations take  part  in  the  one-loop
effects  through  the  mixing  between  the  mass eigenstates and  the  weak
eigenstates of the quark wavefunctions [11]. This means that  the
physics  of  $K^0-\bar K^0$, $B_d^0-\bar B_d^0$ and $B_s^0-\bar B_s^0$ mixing
will be affected  by  $\chi$,  and  the  same  is  true  for  the
$CP$-violating  $\epsilon$  parameter. In an earlier paper [8] we
have discussed these effects in detail and  showed  that  we  can
obtain   a  lower  bound  on  the  mass  of  the  charged  scalar
$\chi^{+}$, which is of the order  of  a  few  hundreds  of  GeV.
Another  bound  can  be  extracted from  the observed rate of the radiative
$B$-decays [12], which is of the same  order  of  magnitude,  and
which is free from a number of undetermined or  poorly
determined parameters which entered in ref. 8. We have also shown
that  the maximum mass splitting in the doublet cannot be greater
than 47 GeV [13]. This last result is obtained from the present experimental
bounds on the oblique electroweak parameters [14].

The chief obstacle to putting phenomenological constraints on the
model  from  low-energy  data  such  as  $B_d^0-\bar  {B_d^0}$ mixing
arises, as usual, from uncertainties in the  hadronic  parameters
such  as  the decay constants $f_K$, $f_B$ and the bag parameters
$B_K$, $B_B$. In this note we investigate a different  observable,
{\it viz}, the ratio $R_b$, defined as
$$
R_b={\zbb\over {\Gamma (Z\r \ {\rm hadrons)}}}. \eqno(6)
$$
$R_b$   is   relatively   free  from  uncertainties  in  hadronic
parameters which tend to cancel out  of  numerator  and
denominator.  It  is also relatively insensitive to $m_t$ and QCD
corrections. For this reason, the effects of new physics can show
up  in  $R_b$  {\it  {without  being masked}} by uncertainties in
$m_t$  etc.,  as  in  the   case   with   a   number   of   other
phenomenologically  interesting  parameters.  An  analysis of the model using
$R_b$ is also facilitated by the fact that the experimental error
in its determination has come down drastically with the LEP
measurements and the advent of
microvertex  detectors, and now stands at [15]
$$R_b\ (expt.)=0.2201\pm 0.0031\eqno(7)$$
at $95\%$ C.L., which is remarkably precise.

To fix ideas and notations, let us briefly discuss the features of
$\zbb$ and $R_b$ in the  Standard  Model  [16-17].  The
tree-level contribution to $\zbb$ is
$$
\zb={G_{\mu}m_Z^3\over  {8\pi\surd  2}}\sqrt{1-4\mu_b}  \Big  [
1-4\mu_b+(1-\textstyle {4\over 3}\sin ^2\theta_W )^2(1+2\mu_b) \Big
] \eqno(8)
$$
where  $\mu_b=m_b^2/m_Z^2$,  $G_{\mu}$ is the Fermi coupling constant
as obtained from muon decay
and $\theta _W$ is the weak mixing angle.

The  electroweak  radiative  corrections  appear  in  the  form  of  two
form factors $\kappa_b$ and $\rho_b$, respectively  for  effective
mixing angle and the overall  renormalization.  Thus,  the  decay
width, calculated to one-loop, is given by
$$
\z={G_{\mu}m_Z^3\over  {8\pi\surd  2}}\rho_b
\sqrt{1-4\mu_b}  \big  [ 1-4\mu_b+(1-\textstyle {4\over 3}
\sin ^2\theta_W \kappa_b)^2(1+2\mu_b) \Big ] \eqno(9)
$$
where $\sin ^2\theta_W$ is determined from
$$\sin  ^2\theta_W  \cos  ^2  \theta  _W={\pi\alpha\over {\surd 2
G_{\mu}m_Z^2(1-\Delta r)}}, \eqno(10)$$
$\Delta r$ being the electroweak correction to $\mu ^{\pm}$ decay.

The   one-loop   correction   is   dominated  by  the  top  quark
contribution. The vacuum polarization effect, which is common to all
fermionic final states, is denoted by $\Delta\rho_t$, which is given by
$$
\Delta\rho _t={\Pi_{ZZ}(0)\over      m_Z^2}-{\Pi_{WW}(0)\over
m_W^2}={3G_{\mu}m_t^2\over  {8\pi^2\surd  2}}\approx {\alpha\over
\pi}{m_t^2\over m_Z^2}, \eqno(11)
$$
the $b$-mass having been neglected.  The  $\Pi$  functions  are  the
standard ones used to denote the vacuum polarization of the gauge
bosons. For  $Z\r  b\bar  b$,  the  vertex
corrections give
$$\Delta\rho _b=-{4\over 3}\Delta\rho _t\eqno(12a)$$
$$\ssq\Delta\kappa _b={2\over 3}\ssq\Delta\rho _t. \eqno(12b)$$
Taking both these factors into account, we can write
$$\rho_b=1+\Delta\rho _t+\Delta\rho _b+\cdots \eqno(13a)$$
$$\kappa_b \ssq=\ssq+\csq\Delta\rho  _t+{2\over 3}\ssq\Delta\rho _t+
\cdots \eqno(13b)$$
After some straightforward algebra, it can be shown that
$$R_b={13\over  59}\Big ( 1+{46\over 59}\bl +{24\over 767}{g_{Vl}
\over g_{Al}}+{0.1\alpha_s (m_Z^2)\over \pi}\Big )\eqno(14)$$
with $\ssq=0.2324$, and $\bl$ and $g_{Vl}/g_{Al}$ are given by
$$\bl\simeq     -{20\alpha\over     13\pi}{m_t^2\over      m_Z^2}
-{10\alpha\over 3\pi}\ln {m_t^2\over m_Z^2} \eqno(15)$$
$${g_{Vl}\over g_{Al}}=1-4 \ssq. \eqno(16)$$
The top-dependent contribution is  of  the  order  of  $10^{-3}$,
the  same  as  the  order  of  experimental  errors, and thus the
variation of $R_b$ with $m_t$  over  the  allowed  range  of  the
latter  is  rather  flat. As stated above, this is one of the reasons
for which $R_b$ is phenomenologically interesting.
In our model, another term of the form
$$\delta   ^{\chi}(R_b)={13\over 59} {46\over 59}(\del   )
\eqno(17)$$
gets added to the above contribution. We take only the  non-oblique
part  as  it  is known [13] that the oblique part has negligible
contribution.
It  is  noteworthy  that the effective Lagrangian only
favors the production of left-handed $b$ quarks,  but  since  the
same  is  true  for  the  tree-level  case, it will not cause any
significant change in the electroweak asymmetries.

In the limit $m_b\r 0$, we can introduce the effects of  the  new
physics  through  a change in the vertex factors for the
$Z\rightarrow b\bar b$ coupling:
$$v'_L=v_L+{8\over 3}\coeff F_L(P^2,m_t) \eqno(18a)$$
$$v'_R=v_R\eqno(18b)$$
where $P$ is the four-momentum of the external $Z$, and
the  color factor of 8/3 comes from the octet nature of $\chi$
under $SU(3)_c$. The right-handed coupling is  not  changed  as  no
term of the form $\bar t_L b_R\chi$ is allowed in the Lagrangian.
The function  $F_L$  represents  the  total of all one-loop  correction
effects,  depicted  in  Fig.  1.  It can be written as the sum of
three terms,
$$F_L=F_L^a+F_L^b+F_L^c,\eqno(19)$$
where $F_L^a$, $F_L^b$ and $F_L^c$ denote the contributions  from
the  figures 1a, 1b and 1c respectively. The correction works out
to be
$$\del={8\over 3}\coeff {2v_L\over {v_L^2+v_R^2}} F_L(m_Z^2, m_t)
\eqno(20)$$
with
$$v_L=-{1\over 2}+{1\over 3}\ssq,\ \ v_R={1\over 3}\ssq.\eqno(21)$$
The $F_L$ functions are
$$F_L^a=b_1(m_{\chi},m_t,m_b^2)v_L\lambda_L^2,\eqno(22a)$$
$$F_L^b=\Bigg  [\Big  [  {P^2\over  \mu_R^2}c_6(m_{\chi},m_t,m_t)
-{1\over 2}-c_0(m_{\chi},m_t,m_t)\Big ] v_R^{(t)}$$
$$\  \ +{m_t^2\over \mu_R^2}c_2(m_{\chi},m_t,m_t)v_L^{(t)}\Bigg ]
\lambda_L^2, \eqno(22b)$$
$$F_L^c=c_0(m_t,m_{\chi},m_{\chi})(\textstyle{1\over      2}-s^2)
\lambda_L^2 \eqno(22c)$$
where  $\lambda_L=g'_t/g$, $g$ being the usual $SU(2)_L$ coupling
constant, $m_{\chi}$ is the mass of the charged  $\chi$,  $\mu_R$
is the mass scale arising in dimensional regularization, and
$$v_R^{(t)}=-{2\over  3}\ssq,\  \  v_L^{(t)}={1\over 2}-{2\over 3}
\ssq. \eqno (23)$$
The two- and three-point functions $b_1$, $c_0$, $c_2$ and $c_6$ in
terms  of the well-known Passarino-Veltman functions [18] are
[16]
$$b_1(m_1,m_2)=B_1(m_2,m_1)+{1\over 2}(\Delta-\ln  \mu_R^2),\eqno
(24a)$$
$$c_0(m_1,m_2,m_3)=-2C_{24}(m_2,m_1,m_3)+{1\over    2}(\Delta-\ln
\mu_R^2), \eqno(24b)$$
$$c_2(m_1,m_2,m_3)=\mu_R^2 C_0(m_2,m_1,m_3), \eqno(24c)$$
$$c_6(m_1,m_2,m_3)=-\mu_R^2    [C_{23}+C_{11}]     (m_2,m_1,m_3)
\eqno(24d)$$
where   $\Delta=2/(4-d)-\gamma-\ln\pi$ in $d$ dimensions, and this
divergence cancels out in the final formula for $F_L$.

In Fig. 2 we show the plot of $\delta ^{\chi}(R_b)$ with $m_{\chi}$
for the top mass ranging from 110 GeV to 200 GeV.
The corresponding $g'_t$ values can be obtained from eq. (4). We have
taken $\ssq=0.2324$, $m_b=4.7$ GeV and $\alpha_s (m_Z^2)=0.117$. We
have checked that very little change in the final results
occur if we take into account the errors in $\alpha_s (m_Z^2)$
and $\ssq$.

In Fig. 3 we plot the lower bound on $m_{\chi}$ for different $m_t$,
ranging from 100 to 180 GeV. It may be noted that this bound goes as $m_t^2$
and $m_t=190$ GeV is the maximum allowed limit.
For $m_t=150$ GeV, we get $m_{\chi}
=380$ GeV as the lower limit. It is
to  be  noted  that the $\chi$-contribution, being negative, makes
the bound very stringent in  nature. The  new  physics
decouples  in  the  limit $m_{\chi}\r
\infty$; this result is in conformity with those obtained earlier
[8,  12-13].  Of  course,  this is just a technical point since $\chi$ is a
composite object and it is meaningless to carry $m_{\chi}$ beyond
the   compositeness  scale.  So it can be claimed that {\it {within the
framework of this model}}, $m_{\chi}$ has
both an upper as well as a lower bound, and these come closer  and
finally  coincide  as  the  cutoff  $\Lambda$  is decreased. This
behaviour can be easily  explained;  if  we  decrease  $\Lambda$,
$(m_t)_{BHL}$  will  increase,  so  there will be a corresponding
increase  in  $g'_t$,   and   $\delta_b^{\chi}$   functions   are
proportional  to  ${g'_t}^2$.  The  coincidence  occurs  at about
$\Lambda=1$ TeV.

In this work, therefore, we  have  investigated  the  effects  of
isodoublet  color-octet  composite scalars arising in a realistic
model  with  dynamical  breaking  of  electroweak  symmetry.  The
specific  process focussed on is the decay $Z\rightarrow b\bar b$, since the
ratio  $R_b$  is  precisely  determined  and  well-known  to   be
relatively   free   from  uncertainties  in  the  Standard  Model
parameters such as $m_t$. We find that a stringent lower bound  can  be
placed on the masses of these composite colored scalars, which is
around 400  GeV  for  a  top  mass  of  150  GeV,  and  increases
quadratically with $m_t$. At the present moment, one
of the main issues confronting $Z$ decay  experiments  is  to
determine the difference between the Standard Model prediction of
$R_b$ and the experimental  number,  because  this  is  one  more
possible gateway to look into the new physics. The Minimal Supersymmetric
Standard Model predicts a positive (but small) $\delta R_b$,  and
nearly all extensions and modifications
in the scalar sector (whether elementary or
composite) predict $\delta R_b$ to be slightly negative. None of
the  alternatives  can be ruled out at the present moment, and
further precision  experiments  could  help discriminate between
models.

\vskip 1 true cm

\c {\bf {Acknowledgements}}

We are indebted to Gautam Bhattacharyya for valuable discussions.
The  two- and three-point functions were calculated using the code
CN developed by Biswarup Mukhopadhyaya and Amitava Raychaudhuri.

\vfill\eject

\c {\bf {References}}

\item {1.} Y. Nambu, in {\it {Proc.  of  the  1988  International
Workshop  on  New  Trends  in  Strong  Coupling Gauge Theories}},
Nagoya, Japan, eds. by M. Bando, T. Muta and K.  Yamawaki  (World
Scientific,  Singapore,  1989);  Y.  Nambu,  EFI Report No. 89-08
(1989), (unpublished).

\item {2.} W.A. Bardeen, C.T. Hill and  M.  Lindner,  Phys.  Rev.
{\bf D41}, 1647 (1990).

\item  {3.}  V.A.  Miransky,  M. Tanabashi and K. Yamawaki, Phys.
Lett. {\bf B221}, 177 (1989); Mod. Phys.  Lett.  {\bf  A4},  1043
(1990).

\item {4.} W.J. Marciano, Phys. Rev. {\bf D41}, 219 (1990).

\item {5.} M. Tanabashi, in {\it {Proc. of the 1991 Nagoya Spring School
on Dynamical Symmetry Breaking}}, Nagoya, Japan, ed. by K. Yamawaki
(World Scientific, Singapore, 1992).

\item  {6.}  Y.-b. Dai {\it {et al}}, Phys. Lett. {\bf B285}, 245
(1992).

\item {7.} A. Kundu, T. De and B.  Dutta-Roy,  Mod.  Phys.  Lett.
{\bf A8}, 2465 (1993).

\item  {8.}  A.  Kundu,  T.  De  and B. Dutta-Roy, Saha Institute
Report No. SINP-TNP/93-12 (to appear in Phys. Rev. {\bf D}).

\item {9.} S. Okubo, Prog. Theor. Phys. {\bf 27}, 949 (1962);  A.
Kundu,  T.  De  and  B.  Dutta-Roy,  Saha  Institute  Report  No.
SINP-TNP/93-08 (to appear in J. Group Theory in Physics).

\item {10.} T. Kugo, in {\it {Proc. of the 1991 Nagoya Spring School
on Dynamical Symmetry Breaking}}, Nagoya, Japan, ed. by K. Yamawaki
(World Scientific, Singapore, 1992).

\item {11.} M. Kobayashi and T. Maskawa, Prog. Theor. Phys.  {\bf
49}, 652 (1973).

\item  {12.}  A.  Kundu,  T.  De and B. Dutta-Roy, Saha Institute
Report No. SINP-TNP/93-15 (to appear in Phys. Rev. {\bf D}).

\item  {13.}  G.  Bhattacharyya  {\it  {et  al}},  University  of
Calcutta Report No. CUPP-93/6 (communicated).

\item  {14.}  J.  Ellis, G.L. Fogli and E. Lisi, Phys. Lett. {\bf
B285}, 238 (1992), {\it ibid} {\bf B292}, 427 (1992).

\item {15.} A. Blondel  and  C.  Verzegnassi,  Phys.  Lett.  {\bf
B311}, 346 (1993).

\item  {16.}  M.  Boulware  and D. Finnell, Phys. Rev. {\bf D44},
2054 (1991).

\item {17.} A.A. Akhundov, D.Yu. Bardin  and  T.  Riemann,  Nucl.
Phys.  {\bf B276}, 1 (1986); W. Beenakker and W. Hollik, Z. Phys.
{\bf C40}, 141 (1988); J. Bernabeu, A. Pich  and  A.  Santamaria,
Phys.  Lett.  {\bf B200}, 569 (1988); F. Boudjema, A. Djouadi and
C. Verzegnassi, Phys. Lett. {\bf B238},  423  (1990);  A.  Denner
{\it  {et  al}},  Z. Phys. {\bf C51}, 695 (1991); A. Djouadi {\it
{et al}}, Nucl. Phys. {\bf B349}, 48 (1991).

\item {18.} G. 't Hooft and M. Veltman, Nucl. Phys. {\bf B153}, 365 (1979);
G. Passarino and M. Veltman, Nucl. Phys. {\bf  B160},
151 (1979).
\vfill\eject

\noindent {\bf {Figure Captions}}

\item  {1.}   The  one-loop diagrams involving the colored scalars
$\chi ^{\pm}$
which contribute to $\zbb$.

\item {2.} The contribution of the $\chi ^{\pm}$ to  the  parameter
$R_b$ as a function of $m_{\chi ^{+}}$. The shaded region depicts
the phenomenologically allowed region for the top mass.

\item {3.} The lower bound $\bar m_{\chi}$ on the mass of ${\chi ^{\pm}}$
 as  a  function  of
$m_t$ for $R_b=0.2170$.

\vfill\eject\end